\documentclass[aps,twocolumn,showpacs,superscriptaddress]{revtex4-1}
\usepackage{times}
\usepackage{graphicx}
\usepackage{amsmath}
\usepackage{chemarrow}
\usepackage{color}
\newcommand{\bra}[1]{\langle #1|}
\newcommand{\ket}[1]{| #1 \rangle}  
\def\degree{{}^{\circ}}

\begin{document}
\title{Quantum Image Processing and Its Application to Edge Detection: Theory and Experiment}
\author{Xi-Wei Yao}
\email{yau@xmu.edu.cn}
\address{College of Physical Science and Technology, Xiamen University, Xiamen, Fujian 361005, China}

\author{Hengyan Wang}
\address{CAS Key Laboratory of Microscale Magnetic Resonance and Department of Modern Physics, University of Science and Technology of China, Hefei, Anhui 230026, China}
\address{Department of Physics, Zhejiang University of Science and Technology, Hangzhou, Zhejiang 310023, China}

\author{Zeyang Liao}
\address{Institute for Quantum Science and Engineering (IQSE) and Department of Physics and Astronomy, Texas A$\&$M University, College Station, Texas 77843-4242, USA}

\author{Ming-Cheng Chen}
\address{Hefei National Laboratory for Physical Sciences at Microscale and Department of Modern Physics, University of Science and Technology of China, Hefei, Anhui 230026, China}

\author{Jian Pan}
\address{CAS Key Laboratory of Microscale Magnetic Resonance and Department of Modern Physics, University of Science and Technology of China, Hefei, Anhui 230026, China}

\author{Jun Li}
\address{Beijing Computational Science Research Center, Beijing  100193, China}

\author{Kechao Zhang}
\address{Institute of Theoretical Physics, Chinese Academy of Sciences, Beijing 100190, China}

\author{Xingcheng Lin}
\address{Department of Physics $\&$ Astronomy and Center for Theoretical Biological Physics, Rice University, Houston, Texas 77005, USA}

\author{Zhehui Wang}
\address{School of Mathematical Sciences, Peking University, Beijing 100871, China}

\author{Zhihuang Luo}
\address{Beijing Computational Science Research Center, Beijing  100193, China}

\author{Wenqiang Zheng}
\address{Center for Optics and Optoelectronics Research, College of Science, Zhejiang University of Technology, Hangzhou, Zhejiang 310023, China}

\author{Jianzhong Li}
\address{College of mathematics and statistics, Hanshan Normal University, Chaozhou, Guangdong 521041, China}

\author{Meisheng Zhao}
\address{Shandong Institute of Quantum Science and Technology, Co., Ltd., Jinan, Shandong 250101, China}

\author{Xinhua Peng}
\email{xhpeng@ustc.edu.cn}
\address{CAS Key Laboratory of Microscale Magnetic Resonance and Department of Modern Physics, University of Science and Technology of China, Hefei, Anhui 230026, China}
\address{Synergetic Innovation Center of Quantum Information and Quantum Physics, University of Science and Technology of China, Hefei, Anhui 230026, China}

\author{Dieter Suter}
\email{dieter.suter@tu-dortmund.de}
\address{Fakult\"{a}t Physik, Technische Universit\"{a}t Dortmund, D-44221 Dortmund, Germany}

\begin{abstract}
Processing of digital images is continuously gaining in volume and relevance, with concomitant demands on data storage, transmission and processing power.
Encoding the image information in quantum-mechanical systems instead of classical ones and replacing classical with quantum information processing may alleviate some of these challenges.
By encoding and processing the image information in quantum-mechanical systems, we here demonstrate the framework of quantum image processing, where a pure quantum state encodes the image information: we encode the pixel values in the probability amplitudes and the pixel positions in the computational basis states.
Our quantum image representation reduces the required number of qubits compared to existing implementations, and we present image processing algorithms that provide exponential speed-up over their classical counterparts.
For the commonly used task of detecting the edge of an image, we propose and implement a quantum algorithm that completes the task with only one single-qubit operation, independent of the size of the image. This demonstrates the potential of quantum image processing for highly efficient image and video processing in the big data era.
\end{abstract}

\pacs{ 03.67.Ac, 07.05.Pj, 32.30.Dx}
\maketitle

\section{Introduction}\label{sec:Intro}
Vision is by far the most important channel for  obtaining information.
Accordingly, the analysis of visual information is one of the most important functions of the human brain \cite{Marr1982}.
In 1950, Turing proposed the development of machines that would be able to ``think'', i.e. learn from experience and draw conclusions, in analogy to the human brain.
Today, this field of research is known as artificial intelligence (AI) \cite{Turing1950,AlphaGo2016,Learning2014}.
Since then, the analysis of visual information by electronic devices has become a reality that enables machines to directly process and analyze the information contained in images and stereograms, or video streams, resulting in rapidly expanding applications in widely separated fields like biomedicine, economics, entertainment, and industry (e.g., automatic pilot) \cite{GonzalezDIPBook,Lake2015,Jean2016}.
Some of these tasks can be performed very efficiently by digital data processors, but others remain time-consuming.
In particular, the rapidly increasing volume of image data as well as increasingly challenging computational tasks have become important driving forces for further improving the efficiency of image processing and analysis.

Quantum information processing (QIP), which exploits  quantum-mechanical phenomena such as quantum superpositions and quantum entanglement \cite{Deutsch1985,Knill2001,Knill2005,Browne2005,MajoranaQC2016,Pan2012RMP,SolidGeoGate2014,Teleport2015,QIQM2015Zeng,Chuang2016,Suter2016RMP,ZhangJ2015,EntanglNNS2017,Micius2017,Liao2017,Ren2017},
allows one to overcome the limitations of classical computation and reaches higher computational speed for certain problems like factoring large numbers \cite{Shor1994,Peng2008} ,
searching an unsorted database \cite{Grover1997},
boson sampling \cite{Aaronson2011,Broome2013,Spring2013,Tillman2013,Crespi2013,LuPan2017Boson},
quantum simulation \cite{Lloyd1996,Peng2009,Peng2010,AD2010,Peng2014,LocdelocTrs2015,CMC2016,LiOTOC2017}, solving linear systems of equations \cite{HHL2009,Cai2013,PJ2014,Barz2014,SupCond2017Lineq},  and machine learning \cite{QMLearn2015Lu,Li2015,QMLearn2017SupCond}.
These unique quantum properties, such as quantum superposition and quantum parallelism, may also be used to speed up signal and data processing \cite{WBL2012,PCA2014}.
For quantum image processing, quantum image representation (QImR) plays a key role, which substantively determines the kinds of processing tasks and how well they can be performed. A number of QImRs \cite{QImR2016,Venegas2003,Le2011,Z2013} have been discussed.

In this article,  we demonstrate the basic framework of quantum image processing based on a different type of QImR, which reduces the qubit resources  required for encoding an image.
Based on this QImR, we experimentally implement several commonly used two-dimensional transforms that are common steps in image processing on a quantum computer and demonstrate that they run exponentially faster than their classical counterparts.
In addition, we propose a highly efficient quantum algorithm for detecting the boundary between different regions of a picture: It requires only one single-qubit gate in the processing stage, independent of the size of the picture.
We perform both numerical and experimental demonstrations to prove the validity of our quantum edge detection algorithm.
These results open up the prospect of utilizing quantum parallelism for image processing.

The article is organized as follows. In Sec. \ref{sec:ImageTrans}, we firstly introduce the basic framework of quantum image processing, then present the experimental demonstration for several basic image transforms on a nuclear magnetic resonance (NMR) quantum information processor.
In Sec. \ref{sec:EdgeDetect}, we propose a highly efficient quantum edge detection algorithm, along with the proof-of-principle numerical and experimental demonstrations. Finally, in Sec. \ref{sec:Conclu}, we summarize the results and give a perspective for future work.

\section{Framework of quantum image processing}\label{sec:ImageTrans}

\begin{figure}[htb]
\centerline{
\includegraphics[scale=1.2]{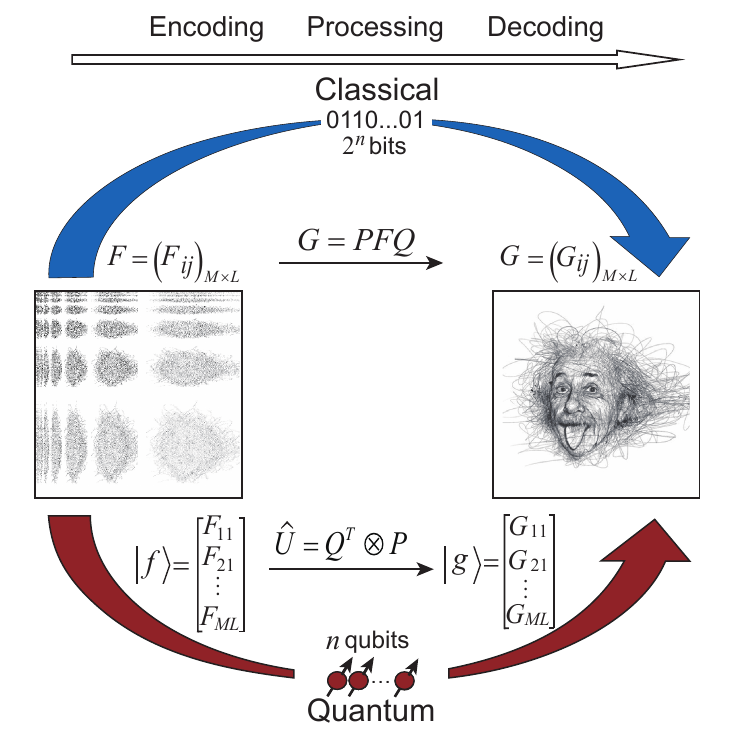}
}
\caption{Comparison of image processing by classical and quantum computers.
$F$ and $G$ are the input and output images, respectively. On the classical computer, an $M \times L$ image can be represented as a matrix and encoded with at least $2^n$ bits [$n=\lceil \log_2 (ML)\rceil$]. The classical image transformation is conducted by matrix computation.  In contrast, the same image can be represented as a quantum state and encoded in $n$ qubits. The quantum image transformation is performed by unitary evolution $\hat{U}$ under a suitable Hamiltonian.
 }
\label{fig:QuanImg}
\end{figure}
In Fig. \ref{fig:QuanImg}, we compare the principles of classical and quantum image processing (QImP). The first step for QImP is the encoding of the 2D image data into a quantum-mechanical system (i.e.,QImR).
The QIR model substantively determines the types of processing tasks and how well they can be performed.
Our present work is based on a QImR where the image is encoded in a pure quantum state, i.e., encoding the pixel values in its probability amplitudes and the pixel positions in the computational basis states of the Hilbert space.
In this section, we  introduce the principle of QImP based on such a QImR, and then present experimental implementations for some basic image transforms, including the 2D Fourier transform, 2D Hadamard, and the 2D Haar wavelet transform.

 \subsection{Quantum image representation}

Given a 2D image $F=(F_{i,j})_{M \times L}$, where $F_{i,j}$ represents the pixel value at position $(i,j)$ with $i = 1,\dots,M$
and $j = 1,\dots,L$, a vector $\vec{f}$ with $ML$ elements can be formed by letting the first $M$ elements of $\vec{f}$ be the first column of $F$, the next $M$ elements the second column, etc. That is,
 \begin{eqnarray}
\vec{f}&=&\text{vec}(F) \nonumber\\
&=& (F_{1,1},F_{2,1},\dots,F_{M,1},F_{1,2},\dots,F_{i,j},\dots,F_{M,L})^T.
\end{eqnarray}
Accordingly, the image data $\vec{f}$ can be mapped onto a pure quantum state $\ket{f} = \sum_{k=0}^{2^{n}-1} c_k \ket{k}$ of $ n=\lceil \log_2 (ML)\rceil$ qubits, where the computational basis $ \ket{k}$ encodes the position $(i,j)$ of each pixel, and the coefficient $c_k$ encodes the pixel value, i.e., $c_k = F_{i,j}/  (\sum {F_{i,j}^2})^{1/2}$ for $k < ML$ and $c_k = 0 $ for $k \geq ML$. Typically, the pixel values must be scaled by a suitable factor before they can be written into the quantum state, such that the resulting quantum state is normalized. When the image data are stored in a quantum random access memory, this mapping takes $O(n) $ steps \cite{QRAM2008}.
 In addition, it was shown that if $c_k$ and $\sum_{k}|c_{k}|^{2}$ can be efficiently calculated by a classical algorithm, constructing  the $n$-qubit image state $\ket{f}$ then takes $O[\text{poly} (n)] $ steps \cite{Grover2002,Soklakov2006}.
Alternatively, QImP could act as a subroutine of a larger quantum algorithm receiving image data from other components \cite{HHL2009}.
Once the image data are in quantum form, they could be postprocessed by various quantum algorithms \cite{Learning2014}.
In Appendix A, we discuss some other QImR models and make a comparison between the QImR we use and others.

\subsection{Quantum image transforms}

Here, we focus on cases where $ML = 2^m \times 2^l$ (an image with $N=ML=2^n$ pixels).
Image processing on a quantum computer corresponds to evolving the quantum state $\ket{f}$ under a suitable Hamiltonian.
A large class of image operations is linear in nature, including unitary transformations, convolutions, and linear filtering (see Appendix C for details).
In the quantum context, the linear transformation can be represented as $\ket{g}=\hat{U} \ket{f}$ with the input image state $\ket{f}$ and the output image state $\ket{g}$.
When a linear transformation is unitary, it can be implemented as a unitary evolution.
Some basic and commonly used image transforms (e.g., the Fourier, Hadamard, and Haar wavelet transforms) can be expressed in the form $G=PFQ$, with the resulting image $G$ and a row (column) transform matrix  $P (Q)$ \cite{GonzalezDIPBook}.
The corresponding unitary operator $\hat{U}$ can then be written as $ \hat{U}={Q}^T \otimes {P}$,
where ${P}$ and ${Q}$ are now unitary operators corresponding to the classical operations. That is, the corresponding unitary operations of $n$ qubits can be represented as a direct product of two independent operations, with one acting on the first $l=\log_2L$ qubits and the other on the last $m=\log_2M$ qubits.

The final stage of QImP is to extract useful information from the processed results.
Clearly, to read out all the components of the image state $\ket{g}$ would require $O( 2^{n}) $ operations.
However, often one is interested not in $\ket{g}$ itself but in some significant statistical characteristics or useful global features about image data \cite{HHL2009}, so it is possibly unnecessary to read out the processed image explicitly.
When the required information is, e.g., a binary result, as in the example of pattern matching and recognition, the number of required operations could be significantly smaller.
For example, the similarity between $\ket{g}$ and the template image $\ket{g'}$ (associated with an inner product $\langle g |g'\rangle$) can be efficiently extracted via the SWAP test \cite{Buhrman2001}  (see Appendix D  for a simple example of recognizing specific patterns).

Basic transforms are commonly used in digital media and signal processing \cite{GonzalezDIPBook}.
 As an example, the discrete cosine transform (DCT), similar to the discrete Fourier transform, is important for numerous applications in science
 and engineering, from data compression of audio (e.g., MP3) and images (e.g., JPEG),
 to spectral methods for the numerical solution of partial differential equations.
High-efficiency video coding (HEVC), also known as H.265,  is one of several  video compression successors to the widely used MPEG-4 (H.264).
Almost all digital videos including HEVC are compressed by using basic image transforms such as 2D DCT or 2D discrete wavelet transforms.
With the increasing amount of data, the running time increases drastically so that real-time processing is infeasible, while quantum image transforms show untapped potential to exponentially speed up over their classical counterparts.

\begin{figure}[htb]
\begin{center}
\includegraphics[scale=1]{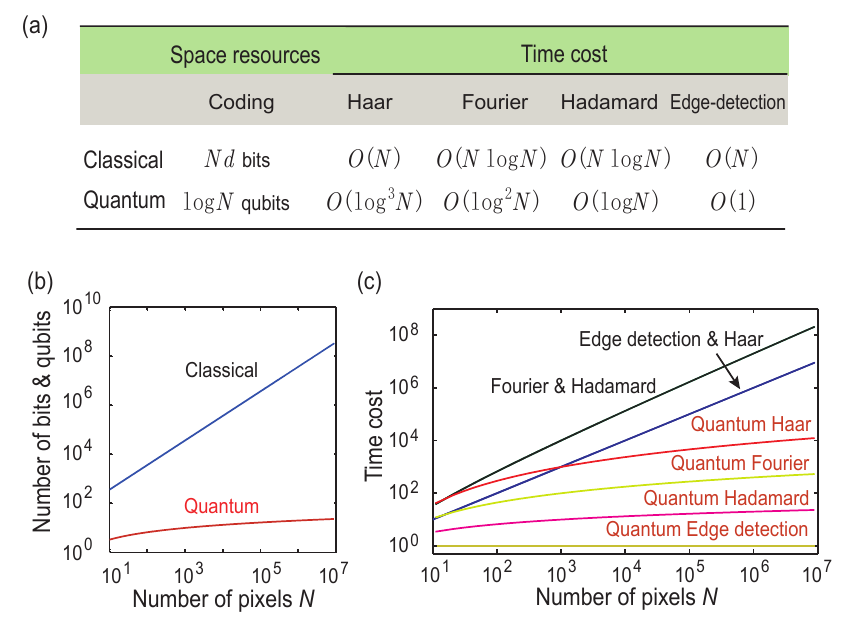}
\end{center}
\caption{
(a) Comparison of resource costs of classical and quantum image processing for an image of $N=M \times L$ (i.e., $n=\log_2 N$) pixels with $d$-bit depth.
(b) Space resources comparison. Top (bottom) curve represents classical (quantum) algorithms, with $d=36$.
(c) Time cost comparison.  The two curves at the top of this graph represent classical algorithms, and the four curves (Quantum Haar, Quantum Fourier, etc.) at the bottom represent quantum algorithms.
 }
 \label{fig:Complexity}
\end{figure}

To illustrate QImP, we now discuss several basic 2D transforms in the framework of QIP, such as the Fourier, Hadamard, and Haar wavelet transforms \cite{QCQI2000Book,Hoyer1997,Fijany1998}.
For these three 2D transforms, $P$ is the transpose of $Q$.
Quantum versions for the one-dimensional Fourier transform (1D QFT) \cite{QFT2001}, 1D Hadamard, and the 1D Haar wavelet transform take time $O[\text{poly}(m)] $, which is polynomial in the number of qubits $m$ (see Appendix B for further details).
However, corresponding classical versions take time $O(m2^{m})$.
When both input data preparation and output information extraction require no greater than $O[\text{poly} (n)] $ steps, QImP, such as the 2D Fourier, Hadamard, and Haar wavelet transforms, can in principle achieve an exponential speed-up over classical algorithms.
 Figure \ref{fig:Complexity} compares the different requirements on resources for the classical and quantum algorithms, in terms of the size of the register (i.e., space) and the number of steps (i.e., time).

\subsection{Experimental demonstrations}
We now proceed to experimentally demonstrate, on a nuclear spin quantum computer, some of these elementary image transforms.
 With established processing techniques \cite{Cory1998,VCS2010}, NMR has been used for many demonstrations of quantum information processing \cite{QFT2001,NC2011Souza,Peng2015,Li2015}.
 \begin{figure}[htb]
\begin{center}
\includegraphics[scale=1.3]{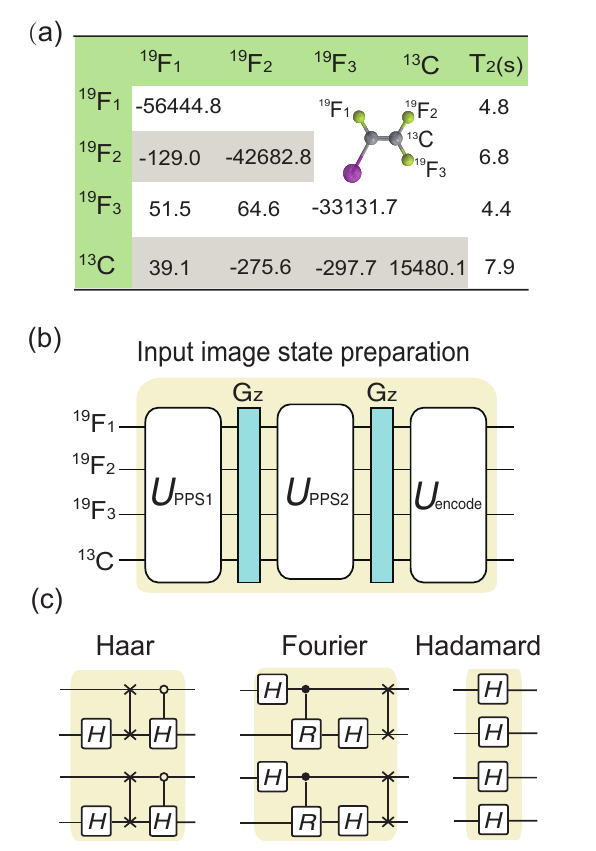}
\end{center}
\caption{(a) Properties of the iodotrifluoroethylene molecule. The chemical shifts and $J$-coupling constants (in Hz) are given by the diagonal and nondiagonal elements, respectively. The measured spin-lattice relaxation times $T_1$ are $21$ s for $^{13}\rm{C}$ and $12.5$ s for $^{19}\rm{F}$. The chemical shifts are given with respect to the reference frequencies of  $100.62$ MHz (carbon)  and $376.48$ MHz (fluorines).
(b) Preparation of the input image states. Two unitary operators $U_{\text{PPS1}}$ and $U_{\text{PPS2}}$ and two $z$-axis gradient field pulses are used to prepare the pseudopure state (PPS) $\rho_{0000}$. Then $U_{\text{encode}}$ realizes quantum image encoding. (c) Quantum circuits for the Haar wavelet, Fourier, and Hadamard image transforms, where $H$ is a Hadamard gate and $R= \left[\begin{matrix}
   1 & 0 \\
   0 & i
 \end{matrix}\right]$ is a phase gate.
 }
 \label{fig:SampleCircuits}
\end{figure}

 As a simple test image, we choose a $4 \times 4$ chessboard pattern
 \begin{equation}
 {F_b}=\frac{1}{2\sqrt{2}}
 \left[\begin{matrix}
1 & 0 & 1 & 0 \\
0 & 1 & 0 & 1 \\
1 & 0 & 1 & 0 \\
0 & 1 & 0 & 1 \\
 \end{matrix}\right],
  \label{eq:InputImg}
\end{equation}
whose encoding and processing require four qubits. We therefore chose iodotrifluoroethylene ($\rm{C}_2\rm{F}_3\rm{I}$) as a 4-qubit quantum register, whose molecular structure and relevant properties are shown in Fig. \ref{fig:SampleCircuits}(a).
We label $^{19}\rm{F} _{1}$, $^{19}\rm{F} _{2}$, $^{19}\rm{F} _{3}$,  and $^{13}\rm{C}$ as the first, second, third, and fourth qubit, respectively.
The natural Hamiltonian of this system in the doubly rotating frame \cite{SpinBook} is
\begin{equation}
H_{{\text{int}} } = \sum\limits_{j = 1}^4 {\pi \nu _j\sigma _z^j}  + \sum\limits_{1 \le j < r \le 4}^4 \frac{\pi }{2}{J_{jr}}\sigma _z^j\sigma _z^r,
\end{equation}
where $\nu_j$ represents the chemical shift of spin $j$, and $J_{jr}$ is the coupling constant between spins $j$ and $r$. The experiments were carried out   at $305$ K on a Bruker AV-$400$ spectrometer in a magnetic field of $9.4$ T.

The input image preparation is illustrated in Fig. \ref{fig:SampleCircuits}(b).
Starting from the thermal equilibrium and using the line-selective method \cite{Peng2001}, we prepare the pseudopure state (PPS) $\rho_{0000}=\epsilon|0000\rangle\langle0000|+[(1-\epsilon)/16]I_{16}$, where $\epsilon\approx 10^{-5}$ is the polarization and $I_{16}$ denotes the $16\times 16$ unit operator. The operator $U_{\text{PPS1}}$ equalizes all populations except that of the state $\ket{0000}$, and a subsequent gradient field pulse destroys all coherences except for the homonuclear zero quantum coherences (ZQC) of the $^{19}\rm{F}$ nuclei.
A specially designed unitary operator $U_{\text{PPS2}}$ is applied to the system and transforms these  remaining ZQC to non-ZQC, which are then eliminated by a second gradient pulse.
The resulting  PPS has a fidelity of $98.4\%$ defined by $|\text{tr}(\rho_{\text{th}}\rho_{\text{expt}})|/[\text{tr}(\rho_{\text{th}}^2)\text{tr}(\rho_{\text{expt}}^2)]^{1/2}$,
where $\rho_{\text{th}}$ and $\rho_{\text{expt}}$ represent the theoretical and experimentally measured density matrices, respectively.
The last operator $U_{\text{encode}}$ turns $|0000\rangle\langle0000|$ into the image state
$\rho_{\text{img}}=\ket{f_{\text{img}}}\bra{f_{\text{img}}}$, which corresponds to the input image.
The three unitary operations $U_{\text{PPS1}}$, $U_{\text{PPS2}}$, and $U_{\text{encode}}$ are all realized by gradient ascent pulse engineering (GRAPE) \cite{Grape2005}, each having theoretical fidelity of about $99.9\%$.

\begin{figure}[htb]
\begin{center}
\includegraphics[scale=0.8]{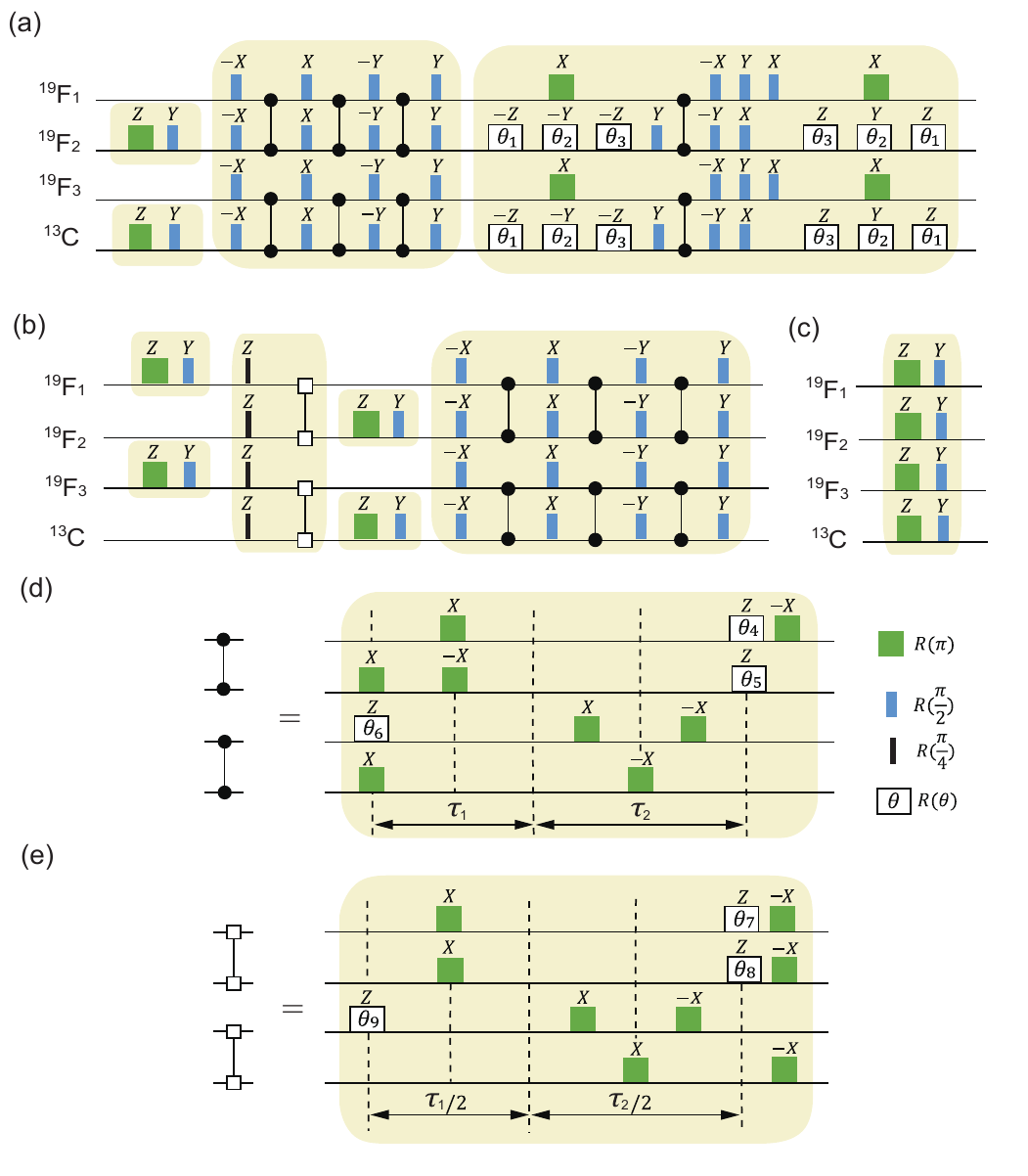}
\end{center}
\caption{ Pulse sequences for  implementing the (a) Haar, (b) Fourier, (c) Hadamard image transform, (d) the operation \rm{exp}$[-i (I_z^1I_z^2+I_z^3I_z^4)\pi]$, and (e) \rm{exp}$\left[i (I_z^1I_z^2+I_z^3I_z^4)\pi/2\right]$.
Here, $\tau_1=|1/2J_{34}|$ and $\tau_2=|1/2J_{12}|-|1/2J_{34}|$, respectively.
The rectangles represent the rotation $R(\theta)$ with the phases given above the rectangles.
The rotation angles $\theta_{1}=-0.1282\pi$, $\theta_{2}=-0.2634\pi$, $\theta_{3}=0.0894\pi$, $\theta_{4}=-2 \pi \nu _1 \tau_2 $, $\theta_{5}=-2 \pi \nu _2 \tau_2 $, $\theta_{6}=-2 \pi \nu _3\tau_1 $, $\theta_{7}=\theta_{4}/2 $,  $\theta_{8}=\theta_{5}/2 $, and $\theta_{9}=\theta_{6}/2 $.
The time order of the pulse sequence is from left to right.
 }
 \label{fig:Pulses}
\end{figure}

For a $4 \times 4$ image, the three image transformation operators that we consider are
\begin{eqnarray}
\hat{U}_{\text{Haar}}=  A_{4}^{ \otimes 2}, \quad \hat{U}_{\text{Fourier}}= \text{QFT}_{4}^{ \otimes 2}, \quad
\hat{U}_{\text{Hadamard}}  =  H^{ \otimes 4},
\end{eqnarray}
 where the Haar, Fourier, and Hadamard matrices are
\begin{equation}
 A_{4}=\frac{1}{2}
 \left[\begin{matrix}
1 & 1 & 1 & 1 \\
1 & 1& -1 & -1 \\
\sqrt{2} & -\sqrt{2} & 0 & 0 \\
0& 0& \sqrt{2} & -\sqrt{2} \\
 \end{matrix}\right],
\end{equation}
\begin{equation}
   \text{QFT}_{4}=\frac{1}{2}
 \left[\begin{matrix}
1 & 1 & 1 & 1 \\
1 & i & -1 & -i \\
1 & -1 & 1 & -1 \\
1 & -i & -1 & i \\
 \end{matrix}\right],
 \end{equation}
 and
 \begin{equation}
H= \frac{1}{\sqrt{2}}\left[\begin{matrix}
   1 & 1 \\
   1 & -1
 \end{matrix}\right].
 \end{equation}
The corresponding quantum circuits and the actual pulse sequences in our experiments are shown in Figs. \ref{fig:SampleCircuits}(c) and \ref{fig:Pulses}, respectively.
 Each unitary rotation in the pulse sequences is implemented through a Gaussian selective soft pulse, and a compilation program is employed to increase the fidelity of the entire selective pulse network \cite{Ryan2008}.
The program systematically adjusts the irradiation frequencies, rotational angles, and transmission phases of the selective pulses, so that up to first-order dynamics, the phase errors and unwanted evolutions of the sequence are largely compensated  \cite{Compiler2016Lijun}.
The resulting fidelities for the $\pi$ refocusing rotations range from $97.2\%$ to $99.5\%$, and for the $\pi/2$ rotations from $99.7\%$ to $99.9\%$. We use the GRAPE technique to further improve the control performance.
The compilation procedure generates a shaped pulse of relatively high fidelity, which serves as a good starting point for the gradient iteration.
So the GRAPE search quickly reaches a high performance.
The final pulse has a numerical fidelity of  $\approx99.9\%$, after taking into account $5\%$ rf  inhomogeneity.
 The whole pulse durations of implementing the Haar, Fourier, and Hadamard transforms are $21.95$, $19.86$, and $3.81$ ms, respectively.

Since the isotropic composition of our sample corresponds to natural abundance, only $\approx 1\%$ of the molecules contain a  $^{13}\rm{C}$ nuclear spin and can therefore be used as quantum registers.
To distinguish their signal from that much larger background of molecules containing $^{12}\rm{C}$ nuclei,
we do not measure the signal of the $^{19}\rm{F}$ nuclear spins directly, but transfer the states of the  $^{19}\rm{F}$ spins to the $^{13}\rm{C}$ spin by a SWAP gate and read out the state information of the  $^{19}\rm{F}$ spins through the $^{13}\rm{C}$ spectra.
Thus, all signals of these four qubits are obtained from the $^{13}\rm{C}$ spectra.

\begin{figure*}[htb]
\centerline{
\includegraphics[scale=1.1]{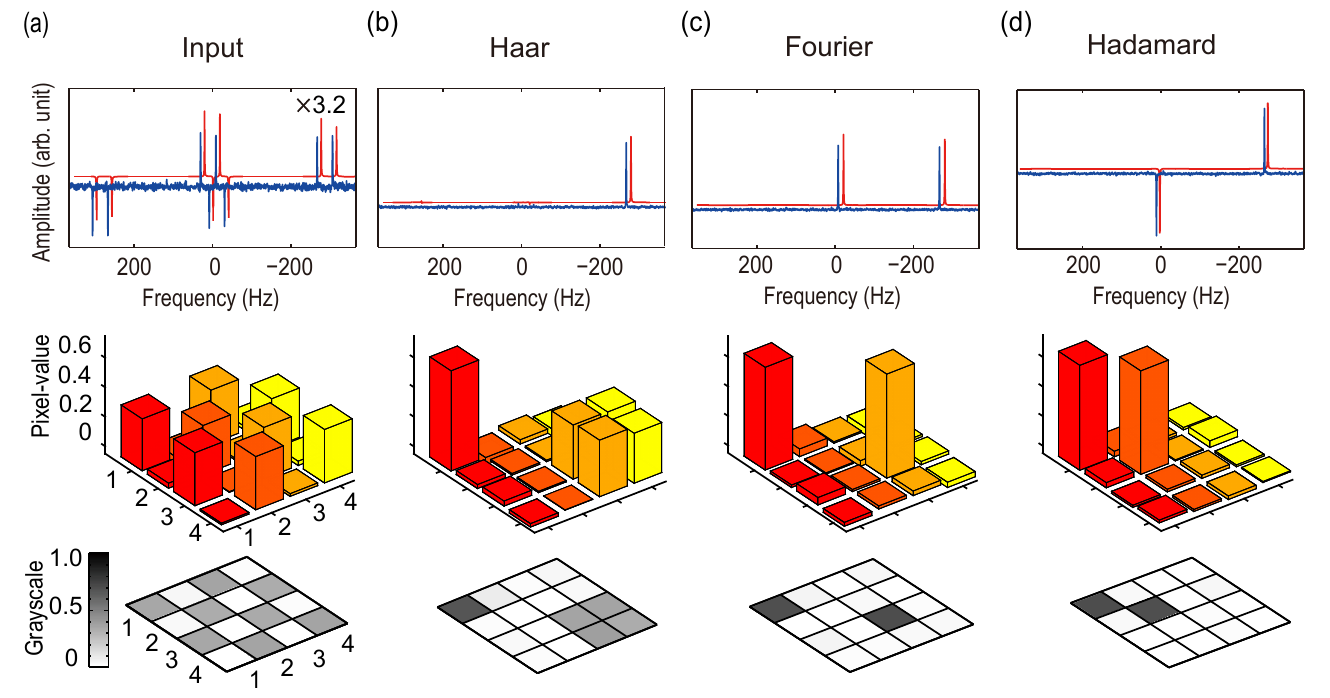}
}
\caption{Experimental results of  quantum image transformations.
(a) Input  $4 \times 4$ image, (b)  Haar-transformed image, (c)  Fourier-transformed image, (d)  Hadamard-transformed image. In (a), the spectral amplitude is zoomed-in by $3.2$ times.
The experimental spectra (top) of the $^{13}\rm{C}$ qubit were obtained by ${\pi}/2$ readout pulses, shown as blue curves. The simulated spectra are denoted as red curves, shifted for clarity.
The experimentally reconstructed images (only real parts are displayed since all imaginary parts are negligibly small) are shown as $3$D bar charts (middle). Their 2D gray scale (visual intensity) pictures (bottom) are displayed with each square representing one pixel in the images.}
\label{fig:Transforms}
\end{figure*}

We apply the Haar wavelet, Fourier, and Hadamard transforms to this input 2D pattern, using the corresponding sequences of rf pulses.
To examine if the experiments have produced the correct results, we perform quantum state tomography \cite{Chuang1998} of the input and output image states.
Compared with theoretical density matrices, the input-image state and the corresponding transformed-image states have fidelities in the range of $[0.961,0.975]$,
As an alternative to quantum state tomography, we also reconstruct  state vectors $\ket{\psi_{\text{expt}}}=\sum_{k=1}^{16}c^{\text{expt}}_k\ket{k}$ directly from the experimental spectra.
The input-image and the transformed-image states are experimentally read out and the decoded image arrays are displayed in  Fig. \ref{fig:Transforms}.  The top row shows the experimental spectra. The middle row shows the corresponding measured image matrices (only the real parts, since the imaginary parts are negligibly small)  as 3D bar charts whose pixel values are equal to the coefficients of the quantum states.
The bottom row represents the same image data as 2D gray scale (visual intensity) pictures.
The experimental and theoretical data agree quite well with each other, with the image Euclidean distance \cite{IMED2009} $\parallel F_{\rm expt} - F_{\rm th}\parallel$/$\parallel F_{\rm th}\parallel \approx   0.08 $ in the input data and  $\in [0.09, 0.12]$ in the resulting data after processing.

\section{Quantum edge detection algorithm}\label{sec:EdgeDetect}

A typical image processing task is the recognition of boundaries (intensity changes) between two adjacent regions \cite{Marr1980Edge}.
This task is not only important for digital image processing, but is also used by the brain:
It has been shown that the brain processes visual information by responding to lines and edges with different neurons \cite{Hubel1995},
which is an essential step in many pattern recognition tasks.
Classically, edge detection methods rely on the computation of image gradients by different types of  filtering masks \cite{GonzalezDIPBook}.
Therefore, all classical algorithms require a computational complexity of at least $O(2^n)$ because each pixel needs to be processed.
A quantum algorithm has been proposed that is supposed to provide an exponential speed-up compared with existing edge extraction algorithms \cite{Qsobel2015}.
However, this algorithm includes a COPY operation and a quantum black box for calculating the gradients of all the pixels simultaneously.
For both steps, no efficient implementations are currently available.
Based on the aforementioned QImR, we propose and implement a highly efficient quantum algorithm that finds the boundaries
between two regions in $O(1)$ time, independent of the image size.
Further discussions regarding more general filtering masks are given in Appendix C.

Basically, a Hadamard gate $H$, which converts a qubit $|0\rangle \to (|0\rangle + |1\rangle)/\sqrt{2}$ and $|1\rangle  \to  (|0\rangle - |1\rangle)/\sqrt{2}$, is applied to detect the boundary.
Since the  positions of any pair of neighboring pixels in a picture column are given by the binary sequences
$b_1 \dots b_{n-1}0 $ and $b_1 \dots b_{n-1}1$, with $b_j= 0$ or $1$, their pixel values are stored as the coefficients
$c_{b_{1}\dots b_{n-1}0}$ and $c_{b_{1} \dots b_{n-1}1}$ of the corresponding computational basis states.
The Hadamard transform on the last qubit changes them to the new coefficients $c_{b_{1}\dots b_{n-1}0} \pm c_{b_{1} \dots b_{n-1}1}$.
The total operation is then
\begin{eqnarray}
 I_{2^{n-1}}\otimes H = \frac{1}{\sqrt{2}}\left[ \begin{array}{ccccccc}
1 & 1 & 0 & 0 & \cdots & 0 & 0 \\
1 & -1 & 0 & 0 & \cdots & 0 & 0\\
0 & 0 & 1 & 1 & \cdots  & 0 & 0\\
0 & 0 & 1 & -1 & \cdots & 0 & 0\\
\vdots & \vdots & \vdots & \vdots & \ddots & \vdots & \vdots\\
0 & 0 & 0 & 0 & \cdots & 1 & 1\\
0 & 0 & 0 & 0 & \cdots & 1 & -1
\end{array} \right],
\end{eqnarray}
where $I_{2^{n-1}}$ is the $2^{n-1}\times 2^{n-1}$ unit matrix.
For an $n$-qubit input image state $\ket{f} = \sum_{k=0}^{N-1} c_k \ket{k}$ ($N=2^{n}$ pixels),
we have the output image state $|g\rangle=\left( I_{2^{n-1}}\otimes H\right)|f\rangle$ as
\begin{eqnarray}\label{eq:HadaEdge}
 I_{2^{n-1}}\otimes H: \left[ \begin{array}{c}c_{0} \\c_{1} \\ c_{2} \\ c_{3}  \\ \cdots \\ c_{N-2} \\ c_{N-1}  \end{array} \right]
\mapsto
\frac{1}{\sqrt{2}} \left[ \begin{array}{c}c_{0}+c_{1} \\ c_{0}-c_{1} \\c_{2}+c_{3} \\ c_{2}-c_{3}  \\ \cdots \\ c_{N-2}+c_{N-1} \\ c_{N-2}-c_{N-1} \end{array} \right].
 \end{eqnarray}
Here, we are interested in the difference $c_{b_{1}\dots b_{n-1}0} - c_{b_{1} \dots b_{n-1}1}$ (the even elements of the resulting state): if the two pixels belong to the same region, their intensity values are identical and the difference vanishes,
otherwise their difference is nonvanishing, which indicates a region boundary.
The edge information in the even positions can be extracted by measuring the last qubit.
Conditioned on the measurement result of the last qubit being $1$, the state of the first $n-1$ qubits encodes the domain boundaries.
Therefore, this procedure yields the horizontal boundaries between pixels at positions 0/1, 2/3, etc.

To obtain also the boundaries between the remaining pairs 1/2, 3/4, etc., we apply the $n$-qubit amplitude permutation to the input image state,
yielding a new image state $|f'\rangle$ with its odd (even) elements equal to the even (odd) elements of the input one $|f\rangle$ (e.g., $c'_{2k} = c_{2k+1}$ and $c'_{2k+1} = c_{2k+2}$). The quantum amplitude permutation can be efficiently performed in $O[poly(n)]$ time \cite{Fijany1998}.
Applying again a single-qubit Hadamard rotation to this new image state $|f'\rangle$, we get the remaining half of the differences.
An alternative approach for obtaining all boundary values is to use an ancilla qubit in the image encoding (see Appendix E for a suitable quantum circuit).
For example, a 2-qubit image state $(c_0, c_1, c_2, c_3)$ can be redundantly encoded in three qubits as $(c_0, c_1, c_1,c_2, c_2,c_3,c_3,c_0)$.
After applying a Hadamard gate to the last qubit of the new image state, we obtain the state $(c_0+c_1, c_0-c_1,c_1+c_2, c_1-c_2,c_2+c_3, c_2-c_3,c_3+c_0, c_3-c_0)$. By measuring the last qubit, conditioned on obtaining $1$, we obtain the reduced state $(c_0-c_1, c_1-c_2,c_2-c_3, c_3-c_0)$, which contains the full boundary information.
With image encoding along different orientations, the corresponding boundaries are detected, e.g., row (column) scanning for the vertical (horizontal) boundary.

This quantum Hadamard edge detection (QHED) algorithm generates a quantum state  encoding the information about the boundary.
Converting that state into classical information will require $O(2^{n})$ measurements, but if the goal is, e.g., to discover if a specific pattern is present in the picture, a measurement of single local observable may be sufficient. A good example is the SWAP test (see Appendix D),
which determines the similarity between the resulting image and a reference image.

\begin{figure}[htb]
\begin{center}
\includegraphics[scale=1]{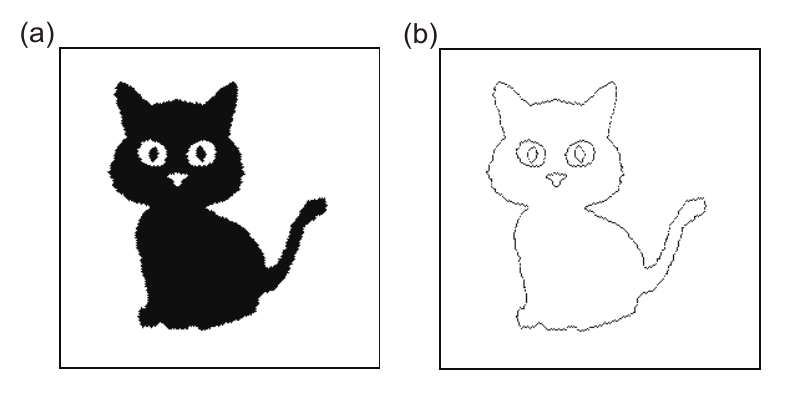}
\end{center}
\caption{Numerical simulation for the QHED algorithm.
(a)  Input $256 \times 256$ image.
(b)  Output image encoding the edge information. The pixels in white and black have  amplitude values $0$ and $1$, respectively.
}
\label{fig:Cat}
\end{figure}
As a numerical example, Fig. \ref{fig:Cat} shows the outcome of the QHED algorithm simulated on a classical computer for an input binary ({\it b}/{\it w}) image $F_{\rm{cat}}$.
For this simple demonstration, we use only a binary image; nevertheless, the QHED algorithm is also valid for an image with general gray levels.
A $256 \times 256$ image $F_{\rm{cat}}$ is encoded into a quantum state $\ket{f_{\rm{cat}}}$ with $16$ qubits instead of $2^{16} =65 536$ classical bits (i.e., $8$ kB). Then a unitary operator $I_{2^{15}} \otimes H$ is applied to $\ket{f_{\rm{cat}}}$.
The resulting image decoded from the output state demonstrates that the QHED algorithm can successfully detect the boundaries in the image.

\begin{figure}[htb]
\begin{center}
\includegraphics[scale=1]{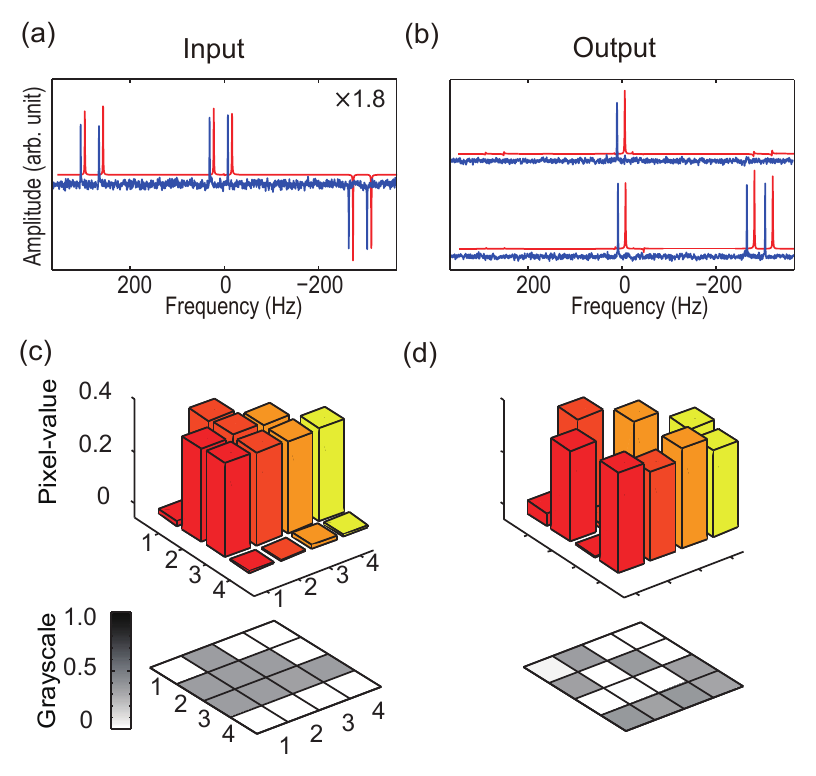}
\end{center}
\caption{Experimental results of the QHED algorithm.
The upper panels are the $^{13}\rm{C}$ spectra (blue curves) for  (a) the input image $F_e$ and (b) output image representing the edge information,
along with the simulated ones (red curves).
The simulated spectra are shifted for clarity. In (a), the spectral amplitude is zoomed-in by $1.8$ times.
In (b), the top (bottom) spectrum is the result after applying a Hadamard gate to $\ket{f_e}$ (the processed image $\ket{f'_e}$ after the amplitude permutation).
The $^{13}\rm{C}$ spectra were obtained by applying ${\pi}/2$ readout pulses.
The lower two panels are the image array results of (c) the input $4 \times 4$ image and (d) the output image representing the edge information. The images are plotted as amplitude $3$D bar charts (top) and 2D visual intensity pictures (bottom) with each square representing one pixel.}
\label{fig:Edge}
\end{figure}

To test the QHED algorithm experimentally, we encode a simple image
 \begin{equation}
  F_e=\frac{1}{2\sqrt{2}}
 \left[\begin{matrix}
0 & 1 & 0 & 0 \\
1 & 1 & 1 & 0 \\
1 & 1 & 1 & 1 \\
0 & 0 & 0 & 0 \\
 \end{matrix}\right]
 \label{eq:InputImgEdge}
\end{equation}
in a quantum state $\ket{f_e}$ of our 4-qubit quantum register.
We then apply  a single-qubit Hadamard gate to the last qubit while keeping the other qubits untouched, i.e.,
$ \hat{U}_e=I_8 \otimes H$.
 The edge information with half of the pixels (even positions) in the resulting state $\ket{g_e} = \hat{U}_e \ket{f_e}$ is produced, which can be read out from the experimental spectra.
We separately perform two experiments to obtain the boundaries for odd and even positions with and without the amplitude permutation, as described above.
To test if the processing result is correct, we measure the input and output image states and obtain their fidelities in the range of $[0.972,0.981]$.
The experimental results of boundary information are shown in Fig. \ref {fig:Edge}, along with some corresponding experimental spectra.
Compared with the theoretical data, the experimental input and output images have image Euclidean distance of $0.06$ and $0.08$, respectively.

\section{Conclusion}\label{sec:Conclu}
In summary, we demonstrate the potential of quantum image processing to alleviate some of the challenges brought by the rapidly increasing amount of image processing.
Instead of the QImR models used in previous theoretical research on QImP, we encode the pixel values of the image in the probability amplitudes and the pixel positions in the computational basis states.
Based on this QImR, which reduces the required qubit resources, we discuss the principle of QImP and experimentally demonstrate the feasibility of a number of fundamental quantum image processing operations, such as the 2D Fourier transform, the Hadamard, and  the Haar wavelet transform, which are usually included as subroutines in more complicated tasks of image processing.
These quantum image transforms provide exponential speed-ups over their classical counterparts.
As an interesting and practical application, we present and experimentally implement a highly efficient quantum algorithm for image edge detection, which employs only one single-qubit Hadamard gate to process the global information (edge) of an image; the processing runs in $O(1)$ time, instead of $O(2^{n})$ as in the classical algorithms.
Therefore, this algorithm has significant advantages over the classical algorithms for large image data.
It is completely general and can be implemented on any general-purpose quantum computer,
such as trapped ions \cite{Ions2011,Ion2017}, superconducting \cite{SuperCond2015,SupCond2017Lineq,QMLearn2017SupCond}, and photonic quantum computing \cite{Kok2007,Carolan2015}.
Our experiment serves as a first experimental study towards practical applications of quantum computers for digital image processing.

In addition to the computational tasks we show in this paper, quantum computers have the potential to  resolve other challenges of image processing and analysis, such as machine learning, linear filtering and convolution, multiscale analysis, face and pattern recognition, image and video coding \cite{Learning2014,QMLearn2015Lu,Li2015,PCA2014,QMLearn2017SupCond}.
Image and video information encoded in qubits can be used not only for efficient processing but also for securely transmitting these data through networks protected by quantum technology.
The theoretical and experimental results we present here may well stimulate further research in these fields.
It is an open area to explore and discover more interesting practical applications involving QImP and AI.
This  paradigm is likely to outperform the classical one and works as an efficient solution in the era of big data.

\begin{acknowledgments}
We are grateful to Emanuel Knill for helpful comments and discussions on the article. We also thank Fu Liu, W. Zhao, X. N. Xu, H. P. Peng, S. Wei, J. Zhang, and X. Y. Zheng for technical assistance, C.-Y. Lu, Z. Chen, S.-Y. Ding, J.-W. Shuai, Y.-F. Chen, Z.-G. Liu, W. Kong, and J. Q. Gu for inspiration and fruitful discussions, and R. Han, X. Zhou, J. Du, and Z. Tian for a great encouragement and helpful conversations.
This work is supported by National Key Basic Research Program of China (Grants No. 2013CB921800 and 2014CB848700), the National Science Fund for Distinguished Young Scholars of China (Grant No. 11425523), the National Natural Science Foundation of China (Grants No. 11375167 and No. 11227901), the Strategic Priority Research Program (B) of the CAS (Grant No. XDB01030400), Key Research Program of Frontier Sciences of the CAS (Grant No. QYZDY-SSW-SLH004), and the Deutsche Forschungsgemeinschaft through Su 192/24-1. Z. Liao acknowledges support from the Qatar National Research Fund (QNRF) under the NPRP Project No. 7-210-1-032. J.-Z. Li. acknowledges support from the Natural Science Foundation of Guangdong Province (Grant No. 2014A030310038). X.-W. Y. and H. W. contributed equally to this work.
\end{acknowledgments}


\section*{Appendix A:  COMPARISON OF QImRs}\label{sec:coding}
Thus far, several QImR models have been proposed.
 In 2003, Venegas-Andraca and Bose suggested the ``qubit lattice'' model to represent quantum images \cite{Venegas2003} where each pixel is represented by a qubit, therefore requiring $2^n$ qubits for an image of $2^n$ pixels. This is a quantum-analog presentation of classical images without any gain from quantum speed-up.
A flexible representation for quantum images (FRQI) \cite{Le2011} integrates the pixel value and position information in an image into an $(n+1)$-qubit quantum state $ (1/{\sqrt{2^{n}}})\sum_{k = 0}^{2^{n} -1} (\cos\theta_k\ket{0}+\sin\theta_k\ket{1})\ket{k}$,
where the angle $\theta_k$ in a single qubit encodes the pixel value of the corresponding position $\ket{k}$.
 A novel enhanced quantum representation (NEQR) \cite{Z2013} uses the basis state $\ket{f(k)}$ of $d$ qubits to store
 the pixel value, instead of an angle encoded in a qubit in FRQI, i.e., an image is encoded as such a quantum state
 $(1/{\sqrt{2^{n}}}) \sum_{k = 0}^{2^{n} -1}  \ket{f(k)}\ket{k}$, where $\ket{f(k)} = \ket{C_k^0 C_k^1\dots C_k^{d-1}}$ with a binary sequence $C_k^0 C_k^1\dots C_k^{d-1}$ encoding the pixel value $f(k)$.
Table \ref{tab:ComparQImR} compares our present QImR, which we refer to as quantum probability image encoding (QPIE), with the other two main quantum representation models: FRQI and NEQR. It clearly shows that the QImR we use here (QPIE) requires fewer resources than the others.

\begin{table*}
\begin{center}
\caption{Comparison of  different QImRs for an image $F= (F_{i,j})_{M \times L}$ with $d$-bit depth  (for the case $M=L=2^m$ and $n = 2m$).
\label{tab:ComparQImR}}
\setlength{\tabcolsep}{7pt}
\renewcommand{\arraystretch}{1.6}
\begin{tabular}{cccc}
\hline
Image representation  & FRQI & NEQR& QPIE \\
\hline
Quantum state & $ (1/{2^{m}})\sum_{k = 0}^{2^{2m} -1} (\cos\theta_k\ket{0}+\sin\theta_k\ket{1})\ket{k}$ & $(1/{2^{m}})\sum_{k = 0}^{2^{2m} -1}  \ket{f(k)}\ket{k}$  & $\ket{f} = \sum_{k=0}^{2^{2m}-1} c_k \ket{k}$\\
Qubit resource           & $1+2m$ & $d+2m$ &$2m$ \\
Pixel-value qubit     &  $1$     & $d$      & $0$ \\
Pixel value           & $\theta_k$ & $f(k) = C_k^0C_k^1\dots C_k^{d-1}$ &$c_k$ \\
Pixel-value encoding  & Angle & Basis of qubits & Probability amplitude \\
\hline
\end{tabular}
\end{center}
\end{table*}

\section*{Appendix B: QUANTUM WAVELET TRANSFORM}\label{sec:Haar}
\setcounter{equation}{0}
\renewcommand\theequation{B\arabic{equation}}
Here, we discuss the  implementation circuit and complexity of the quantum Haar wavelet transform.
Generally, the $M\times M$ Haar  \cite{Haar1910}  wavelet transform  $A_{M}$ ($M=2^m$, $m=0, 1, 2, \dots$) can be defined by the following equation as
\begin{equation}
A_M=\left[\begin{matrix}
A_{M/2} \otimes {B_X}\\
{I_{M/2}} \otimes {B_{\bar{X}}}\\
\end{matrix}\right],
\label{eq:HaarMatrix}
\end{equation}
where  $A_1=1$, $I_{M/2}$ is a $M/2 \times M/2$ unit operator, $B_{X}= \left[ 1 \quad 1\right]/\sqrt{2}$ and
$B_{\bar{X}}= \left[ 1 \quad -1\right]/\sqrt{2}$.
This implies, for $M=2$,
\begin{equation}
A_2=
\left[\begin{matrix}
{B_X}\\
{B_{\bar{X}}}
\end{matrix}\right]
=\frac{1}{\sqrt{2}}
\left[\begin{matrix}
1 & 1\\
1 & -1
\end{matrix}\right]=H,
\label{eq:A2HadamardMatrix}
\end{equation}
 that is, $A_2$ is a Hadamard transform.
 \begin{figure}[ht]
\begin{center}
\includegraphics[scale=0.6]{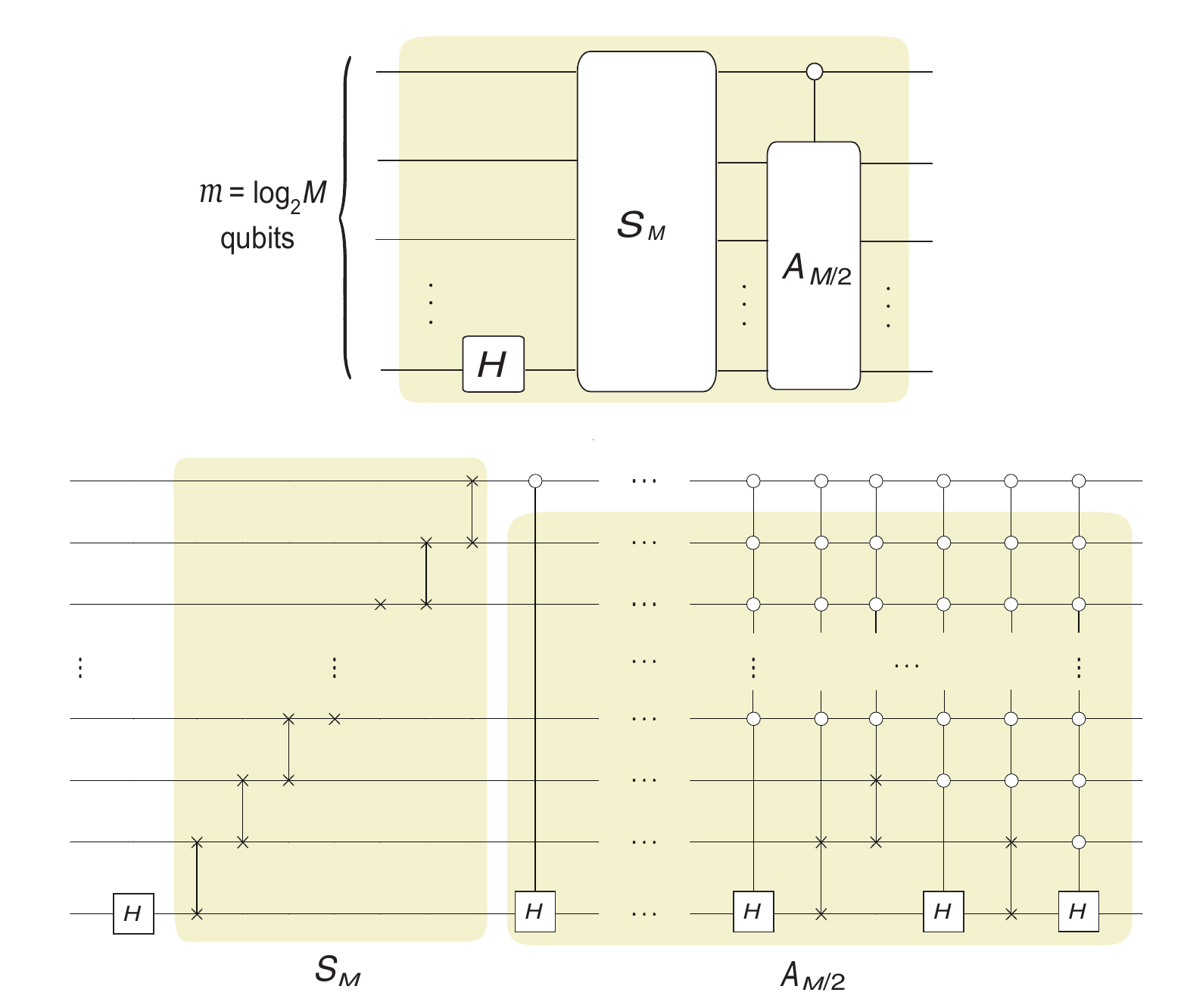}
\end{center}
\caption{Quantum circuit for the Haar wavelet transform $A_{M}$. $H$ is a Hadamard gate, and $A_{2}=H$ for the case $M=2$. $S_{M}$ is the qubit cyclic right shift permutation $S_{M}: \ket{i_1 i_2 \dots i_{m-1} i_m} \rightarrow \ket{i_m i_1 i_2\dots i_{m-1}}$, which can be implemented by $m-1$ $\rm{SWAP}$ gates.  }
 \label{fig:HaarCircuit}
\end{figure}
We can  recursively  decompose  the quantum Haar wavelet transform (Fig. \ref{fig:HaarCircuit}) as follows :
\begin{equation}
A_{M} =
\left[\begin{matrix}
   A_{M/2} & {} \\
   {} & I_{M/2}  \\
   \end{matrix}\right]
S_{M}\left( I_{M/2}\otimes A_2 \right),
\end{equation}
where $S_{M}$ is the qubit cyclic right shift permutation: $S_{M}\ket{i_1 i_2 \dots i_{m-1} i_m}=\ket{i_m i_1 i_2\dots i_{m-1}}$, with $i_j=0$ or $1$ and $m$ the number of qubits. Specifically, $S_4$ is the $\rm{SWAP}$ gate to interchange the states of the two qubits: $ \ket{i_1 i_2} \rightarrow \ket{i_2 i_1}$.
Therefore the corresponding circuit consists of  the following controlled gates.
\begin{enumerate}
\item $C^0(H)$, $C^1(H)$, $C^2(H)$, \dots , $C^{m-1}(H)$,
\item $C^0(S_{2^m})$, $C^1(S_{2^{m-1}})$, \dots, $C^{m-2}(S_4)$.
\end{enumerate}
Here, $C^k(U)$ is a multiple qubit controlled gate described as follows:
\begin{equation}
C^k(U )\ket{i_1 i_2 \dots i_k}\ket{\psi} = \ket{i_1 i_2 \dots i_k}U^{\bar{i}_1\bar{i}_2 \dots \bar{i}_k}\ket{\psi}
\end{equation}
where $\bar{i}_1\bar{i}_2 \dots \bar{i}_k$ in the exponent of $U$ means the product of the bits' inverse $\bar{i}_1\bar{i}_2 \dots \bar{i}_k$, and $\bar{i}={\rm NOT}(i)$.
That is, if the first $k$ control qubits are all in state $\ket{0}$ the $m-k$ qubit unitary operator $U$ is applied to the last $m-k$ target qubits, otherwise the identity operation is applied to the last $m-k$ target qubits.

Since  $S_{2^{m-k}}$ can be implemented by $(m-k-1)$  $\rm{SWAP}$ gates, the circuit for $C^k(S_{2^{m-k}})$ is composed of $(m-k-1)$ $C^k(\rm{SWAP})$ gates. $C^1(\rm{SWAP})$ can be implemented by 3 $C^2(\rm{NOT})$ gates \cite{QCQI2000Book}.
Hence, the  implementation of $C^k(S_{2^{m-k}})$ needs in total $3(m-k-1)$ $C^{k+1}(\rm{NOT})$ gates. Both $C^k(H)$ and $C^k(\rm{NOT})$ can be implemented with linear complexity, for $k=0, \dots, m-1$. Hence, we conclude that the quantum Haar wavelet transform  can be implemented by $O(m^3)$ elementary gates.

\section*{Appendix C: Image spatial filtering}\label{sec:Filter}
\setcounter{equation}{0}
\renewcommand\theequation{C\arabic{equation}}
Spatial filtering is a technique of image processing, such as image smoothing, sharpening, and edge enhancement, by operating the pixels in the neighborhood of the corresponding input pixel. The filtered value of the target pixel is given by a linear combination of the neighborhood pixels with the specific weights determined by the mask values \cite{GonzalezDIPBook}. For example, given an input image $F=(F_{i,j})_{M \times M}$ and a general $3\times 3$ filtering mask,
\begin{equation}\label{eq:mask}
W=\left[\begin{matrix}
w_{11} & w_{12} & w_{13}\\
w_{21} & w_{22} & w_{23}\\
w_{31} & w_{32} & w_{33}\\
\end{matrix}\right],
\end{equation}
spatial filtering will give the output image $G=(G_{i,j})_{M \times M}$ with the pixel $G_{i,j} = \sum_{u,v =1}^{3} w_{uv} F_{i+u-2,j+v-2}$ ($2\leq i, j \leq M-1$).
Here we construct a linear filtering operator $U$ such that  $\vec{g}=U\vec{f}$, where $\vec{f}=\text{vec}(F)$ and $\vec{g}=\text{vec}(G)$.
 $\vec{f}$ and $\vec{g}$ are both $M^2$-dimensional vectors, and the dimension of $U$ is $M^2 \times M^2$.
We prove that $U$ can be constructed as
\begin{equation}\label{eq:U}
U=\left[\begin{matrix}
 E &  & &  & &  &      \\
 V_1 & V_2 & V_3  &  &    &  &   \\
 & \ddots & \ddots & \ddots & & &\\
 &  & \ddots & \ddots &\ddots & &\\
 & & & V_1 & V_2 & V_3\\
   & &  & &  & E    \\
\end{matrix}\right],
\end{equation}
where $E$ is an $M\times M$ identity matrix, and $V_1$, $V_2$, $V_3$ are $M\times M$ matrices defined by
\begin{eqnarray}
V_1=\left[\begin{matrix}
 0 &  & &  & &  &      \\
w_{11} & w_{21} & w_{31}  &&    &  &   \\
 & \ddots & \ddots & \ddots & & &\\
 &  & \ddots & \ddots &\ddots & &\\
 & & & w_{11} &w_{21} & w_{31}\\
   & &  & &  & 0     \\
\end{matrix}\right]_{M\times M},\\
V_2=\left[\begin{matrix}
1 &  & &  & &  &      \\
w_{12} & w_{22} & w_{32}  &&    &  &   \\
 & \ddots & \ddots & \ddots & & &\\
 &  & \ddots & \ddots &\ddots & &\\
 & & & w_{12} &w_{22} & w_{32}\\
   & &  & &  & 1     \\
\end{matrix}\right]_{M\times M},\\
V_3=\left[\begin{matrix}
 0 &  & &  & &  &      \\
w_{13} & w_{23} & w_{33}  &&    &  &   \\
 & \ddots & \ddots & \ddots & & &\\
 &  & \ddots & \ddots &\ddots & &\\
 & & & w_{13} &w_{23} & w_{33}\\
   & &  & &  & 0     \\
\end{matrix}\right]_{M\times M}.
\end{eqnarray}
$ Proof.$-
Since $\vec{g}=\text{vec}(G)$, we have $g_k=G_{t,s+1}$, with $k=t+Ms \quad (1\leq k \leq M^2; 1\leq t \leq M;0\leq s\leq M-1)$. For $t\neq 1,M$ and $s\neq 0,M-1$, we have
\begin{eqnarray*}
 g_k=G_{t,s+1}&=&(W*F)_{t,s+1} \\
&=& w_{11} F_{t-1,s}+w_{21} F_{t,s}+w_{31}F_{t+1,s} \\
& &+w_{12}F_{t-1,s+1}+w_{22} F_{t,s+1}+w_{32} F_{t+1,s+1} \\
& &+w_{13} F_{t-1,s+2}+w_{23} F_{t,s+2}+w_{33} F_{t+1,s+2}
\end{eqnarray*}
Let $\vec{h}= U\vec{f}$, then we have $h_k=\sum\limits_{i=1}\limits^{M^2}U_{k,i}f_i$. From the expression of $U$  in Eq. \eqref{eq:U}, we can see that the nonzero elements are
\begin{eqnarray*}
U_{k,M(s-1)+t-1}=w_{11}, \quad U_{k,M(s-1)+t}=w_{21}, \\
U_{k,M(s-1)+t+1}=w_{31}, \quad U_{k,Ms+t-1}=w_{12}, \\
U_{k,Ms+t}=w_{22},  \quad U_{k,Ms+t+1}=w_{32},\\
U_{k,M(s+1)+t-1}=w_{13}, \quad U_{k,M(s+1)+t}=w_{23}, \\
U_{k,M(s+1)+t+1}=w_{33},
\end{eqnarray*}
 and for other $i$, $U_{k,i}=0$. Since $\vec{f}=\text{vec}(F)$, we have
 \begin{eqnarray*}
f_{M(s-1)+t-1} =F_{t-1,s}, \quad  f_{M(s-1)+t}=F_{t,s},\\
f_{M(s-1)+t+1}=F_{t+1,s}, \quad f_{Ms+t-1}=F_{t-1,s+1}, \\
 f_{Ms+t}=F_{t,s+1}, \quad f_{Ms+t+1}=F_{t+1,s+1}, \\
f_{M(s+1)+t-1}=F_{t-1,s+2}, \quad f_{M(s+1)+t}=F_{t,s+2}, \\
 f_{M(s+1)+t+1}=F_{t+1,s+2}.
\end{eqnarray*}
 By direct comparison, it is readily seen that $h_k=g_k$.
Hence, we have $\vec{g}=U\vec{f}$.

We can deduce that $U$ is unitary if and only if  $w_{22}=\pm1$ and other elements are all zero in $W$ [Eq. \eqref{eq:mask}].
In general,  the linear transformation of spatial filtering is nonunitary.
For a nonunitary linear transformation $U$, we can try to embed it in a bigger quantum system, and perform a bigger unitary operation to realize an embedded  transformation $U$ \cite{Duality2016}.
Alternatively, the quantum matrix-inversion techniques \cite{HHL2009,WBL2012} could also help to perform some nonunitary linear transformations on a quantum computer.

\section*{Appendix D: Detecting symmetry by QImP}\label{sec:InvSymm}
\setcounter{equation}{0}
\renewcommand\theequation{D\arabic{equation}}
Here, we present a highly efficient quantum algorithm for recognizing an inversion-symmetric image, which outperforms state-of-the-art classical algorithms with an exponential speed-up.
First, we use the NOT gate (i.e., the Pauli X operator $\sigma_x$) to rotate the input image $180\degree$  with respect to the image center. Then we utilize the SWAP test \cite{Buhrman2001} to detect the overlap between the input and rotated images: The larger the overlap, the better is the inversion symmetry of original image. This algorithm is described as follows.
\begin{enumerate}
\item Encode an input $M\times L=2^m \times 2^n$ image into a quantum state $\ket{f}$ with $n=m+l$ qubits.
\item Perform a NOT operation on each qubit such that the basis $|i_1 i_2 \dots i_n\rangle$ switches to the complementary basis $|\bar{i}_1 \bar{i}_2 \dots \bar{i}_n\rangle$ (i.e., $U_\text{NOT} =\text{NOT}^{ \otimes n}=\sigma_x^{ \otimes n} $), where $i_{1}, i_{2}, \dots, i_{n}=0$ or $1$ and $\bar{i} = {\rm{NOT}}(i)$. Since $i+\bar{i}=1$, we have $i_{1}i_{2}\dots i_{n}+\bar{i}_1\bar{i}_2\dots\bar{i}_n=11\dots 1$. Therefore, the bases are swapped around the center, i.e., the image is rotated by $180\degree$.
\item Using the SWAP test method \cite{Learning2014,Li2015}, we detect the overlap between two states before and after applying NOT operation to the input pattern; a measured overlap value $\langle f|U_\text{NOT}|f\rangle$ \cite{Knill2007} can efficiently supply useful information on the inversion symmetry of the input pattern.
\end{enumerate}
Estimating distances and inner products between state vectors of image data in $ML$-dimensional vector spaces then takes time $O(\log ML) $ on a quantum computer, which is exponentially faster than that of classical computers \cite{Lloyd2013,Aaronson2009}.
Here a specific example of a $2\times 2$ image is provided for illustration.
To rotate the input image matrix by $180\degree$ as follows,
\begin{equation}
\left[\begin{matrix}
1 & 3\\
2 & 4
\end{matrix}\right]
\autorightarrow{Rotation}{ $180\degree$}
\left[\begin{matrix}
4 & 2\\
3 & 1
\end{matrix}\right].
\end{equation}
The input state of left-hand image is $(|00\rangle+ 2|01\rangle + 3|10\rangle +4 |11\rangle)/\sqrt{30}$.
Applying a NOT gate to each qubit, the input state is transformed to ($|11\rangle+ 2|10\rangle + 3|01\rangle +4 |00\rangle)/\sqrt{30}$ (corresponding to the rotated image on the right-hand side).
It is clear that the input image has been rotated by $180\degree$ around its center, which corresponds to point reflection in 2D.

\section*{Appendix E: Variant of QHED algorithm}\label{sec:AuxQHED}
\setcounter{equation}{0}
\renewcommand\theequation{E\arabic{equation}}

\begin{figure}[h]
\begin{center}
\includegraphics[scale=0.8]{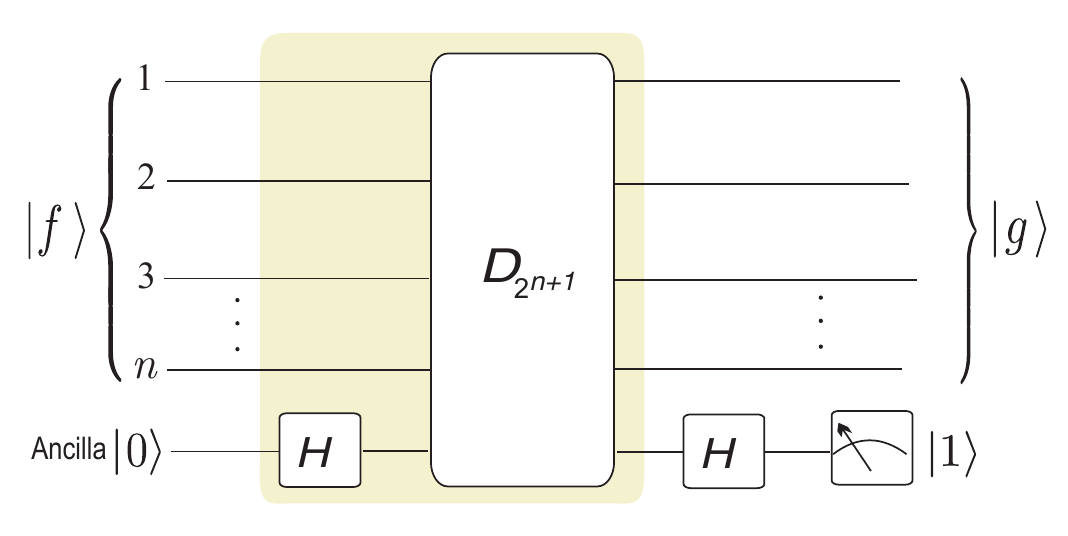}
\end{center}
\caption{Quantum circuit for the QHED algorithm with an auxiliary qubit. $H$ is a Hadamard gate, and $D_{2^{n+1}}$ is an amplitude permutation operation for $n+1$ qubits.
 }
 \label{fig:AuxQHEDCircuit}
\end{figure}

 In order to produce full boundary values in a single step, a variant of the QHED algorithm uses an auxiliary qubit for encoding the image.
 The quantum circuit is shown in Fig. \ref{fig:AuxQHEDCircuit}. The operation $D_{2^{n+1}}$ is an $n+1$-qubit amplitude permutation, which can be written in matrix form as
 \begin{equation}\label{eq:AplPermu}
D_{2^{n+1}} = \left[ \begin{array}{ccccccc}
0 & 1 & 0 & 0 & \cdots & 0 & 0 \\
0 & 0 & 1 & 0 & \cdots & 0 & 0\\
0 & 0 & 0 & 1 & \cdots  & 0 & 0\\
0 & 0 & 0 & 0 & \cdots & 0 & 0\\
\vdots & \vdots & \vdots & \vdots & \ddots & \vdots & \vdots\\
0 & 0 & 0 & 0 & \cdots & 0 & 1\\
1 & 0 & 0 & 0 & \cdots & 0 & 0\\
\end{array} \right].
\end{equation}
It can be efficiently implemented in $O[\text{ploy}(n)]$ time \cite{Fijany1998}.
For an input image encoded in an $n$-qubit state $\ket{f}=(c_0, c_1, c_2,  \dots, c_{N-2},c_{N-1})^T$,
a Hadamard gate is applied to the input state $\ket{0}$ of the auxiliary qubit,
yielding an $(n+1)$-qubit redundant image state $\ket{f}\otimes(\ket{0}+\ket{1})/\sqrt{2}=2^{-1/2}(c_0,c_0,c_1,c_1, c_2, c_2,\dots,c_{N-2},c_{N-2},c_{N-1},c_{N-1})^T$.
The amplitude permutation $D_{2^{n+1}}$ is performed to yield a new redundant image state $2^{-1/2}(c_0,c_1,c_1, c_2, c_2,c_3,\dots,c_{N-2},c_{N-1},c_{N-1},c_0)^T$.
After applying a Hadamard gate to the last qubit of this state, we obtain the state
$2^{-1}(c_0+c_1, c_0-c_1,c_1+c_2,c_1-c_2, c_2+c_3, c_2-c_3,\dots, c_{N-2}+c_{N-1}, c_{N-2}-c_{N-1}, c_{N-1}+c_0, c_{N-1}-c_0)^T$.
By measuring the last qubit, conditioned on obtaining $1$, we obtain the $n$-qubit state  $\ket{g}=2^{-1}(c_0-c_1,c_1-c_2,  c_2-c_3,\dots,c_{N-2}-c_{N-1}, c_{N-1}-c_0)^T$, which contains the full boundary information.

\begin{thebibliography}{99}

\bibitem{Marr1982}
D. Marr, {\it Vision: A Computational Investigation into the Human Representation and Processing of Visual Information}  (W. H. Freeman and Company, San Francisco, 1982).

\bibitem{Turing1950}
A. M. Turing, {\it Computing Machinery and Intelligence}, Mind {\bf 59}, 236, 433-460 (1950).

\bibitem{AlphaGo2016}
D. Silver,	A. Huang, C. Maddison, A. Guez, L. Sifre,	G. Driessche, J. Schrittwieser, I. Antonoglou, V. Panneershelvam, M. Lanctot, S.
Dieleman, D. Grewe, J. Nham, N. Kalchbrenner, L. Sutskever, T. Lillicrap, M. Leach, K. Kavukcuoglu, T. Graepel, and D. Hassabis, {\it Mastering the Game of Go with Deep Neural Networks and Tree Search}, Nature (London) {\bf 529}, 484-489 (2016).

\bibitem{Learning2014}
P. Rebentrost,  M. Mohseni, and S. Lloyd, {\it Quantum Support Vector Machine for Big Data Classification}, Phys. Rev. Lett. {\bf 113}, 130503 (2014).

\bibitem{GonzalezDIPBook}
R. C. Gonzalez and R. E. Woods,  {\it Digital Image Processing} (Prentice-Hall, London, UK, 2007).

\bibitem{Lake2015}
B. M. Lake,  R. Salakhutdinov, and J. B. Tenenbaum, {\it Human-Level Concept Learning through Probabilistic Program Induction}, Science {\bf 350}, 1332-1338 (2015).

\bibitem{Jean2016}
N. Jean, M. Burke, M. Xie, W. M. Davis, D. B. Lobell, and S. Ermon,  {\it Combining Satellite Imagery and Machine Learning to Predict Poverty}, Science {\bf 353}, 790-794 (2016).

\bibitem{Deutsch1985}
D. Deutsch, {\it Quantum Theory, the Church-Turing Principle and the Universal Quantum Computer}, Proc. Royal Soc. London A {\bf 400}, 97-117 (1985).

\bibitem{Knill2001}
E. Knill, R. Laflamme, and G. J. Milburn, {\it A Scheme for Efficient Quantum Computation with Linear Optics}, Nature (London) {\bf 409}, 46-52 (2001).

\bibitem{MajoranaQC2016}
D. Aasen, M. Hell, R. V. Mishmash, A. Higginbotham, J. Danon, M. Leijnse, T. S. Jespersen, J. A. Folk, C. M. Marcus, K. Flensberg, and J. Alicea, {\it  Milestones Toward Majorana-Based Quantum Computing},  Phys. Rev. X {\bf 6}, 031016 (2016).


\bibitem{Micius2017}
J. Yin, Y. Cao, Y. Li, S. Liao, L. Zhang, J. Ren, W. Cai, W. Liu, B. Li, H. Dai, G. Li, Q. Lu, Y. Gong, Y. Xu, S. Li, F. Li, Y. Yin, Z. Jiang, M. Li, J. Jia, G. Ren, D. He, Y. Zhou, X. Zhang, N. Wang, X. Chang, Z. Zhu, N. Liu, Y.-A Chen, C.-Y. Lu, R. Shu, C.-Z. Peng, J.-Y. Wang, and J.-W. Pan, {\it Satellite-Based Entanglement Distribution over 1200 Kilometers}, Science {\bf 356}, 1140-1144 (2017).

\bibitem{Liao2017}
S.-K. Liao et al., Satellite-to-ground Quantum Key Distribution, Nature (London) {\bf 549}, 43-47 (2017).

\bibitem{Ren2017}
J.-G. Ren et al., Ground-to-satellite quantum teleportation, Nature (London) {\bf 549}, 70-73 (2017).

\bibitem{Browne2005}
D. E. Browne and T. Rudolph,  {\it Resource-Efficient Linear Optical Quantum Computation}, Phys. Rev. Lett. {\bf 95}, 010501 (2005).

 \bibitem{Pan2012RMP}
 J.-W. Pan, Z.-B. Chen, C.-Y. Lu, H. Weinfurter,  A. Zeilinger, and M. \.{Z}ukowski, {\it Multiphoton Entanglement and Interferometry}, Rev. Mod. Phys. {\bf 84}, 777-838 (2012).

 \bibitem{Teleport2015}
X. L. Wang, X. D. Cai, Z. E. Su, M.-C. Chen, D. Wu, L. Li, N. L. Liu, C.-Y. Lu, and J.-W. Pan, {\it Quantum Teleportation of Multiple Degrees of Freedom of a Single Photon}, Nature (London) {\bf 518}, 516-519 (2015).

\bibitem{Knill2005}
E. Knill, {\it Quantum Computing with Realistically Noisy Devices}, Nature (London) {\bf 434}, 39-44 (2005).

\bibitem{Suter2016RMP}
D. Suter and G. A. \'{A}lvarez, {\it Protecting Quantum Information against Environmental Noise}, Rev. Mod. Phys. {\bf 88}, 041001 (2016).

\bibitem{ZhangJ2015}
J. Zhang and D. Suter,  {\it Experimental Protection of Two-Qubit Quantum Gates against Environmental Noise by Dynamical Decoupling}, Phys. Rev. Lett. {\bf 115}, 110502 (2015).

 \bibitem{SolidGeoGate2014}
C. Zu, W.-B. Wang, L. He, W.-G. Zhang, C.-Y. Dai, F. Wang, and L.-M. Duan, {\it Experimental Realization of Universal Geometric Quantum Gates with Solid-State Spins}, Nature (London) {\bf 514}, 72-75 (2014).

 \bibitem{QIQM2015Zeng}
B. Zeng, X. Chen, D.-L. Zhou,  and X.-G.Wen,  {\it Quantum Information Meets Quantum Matter -- From Quantum Entanglement to Topological Phase in Many-Body Systems}, arXiv:1508.02595.

\bibitem{Chuang2016}
G. H. Low, T. J. Yoder, and I. L. Chuang, {\it Methodology of Resonant Equiangular Composite Quantum Gates},  Phys. Rev. X {\bf 6}, 041067 (2016).

\bibitem{EntanglNNS2017}
D.-L. Deng, X. Li, and S. D. Sarma, {\it Quantum Entanglement in Neural Network States}, Phys. Rev. X {\bf 7}, 021021 (2017).

\bibitem{Shor1994}
P. W. Shor, Algorithms for Quantum Computation: Discrete Logarithms and Factoring, in {\it Proceedings of the 35th Annual Symposium on Foundation of Computer Science} (IEEE Computer Society Press, New York, 1994), pp. 124-134.

\bibitem{Peng2008}
X.-H. Peng, Z.-Y. Liao, N. Xu, G. Qin, X. Zhou, D. Suter, and J. Du, {\it Quantum Adiabatic Algorithm for Factorization and Its Experimental Implementation}, Phys. Rev. Lett. {\bf 101}, 220405 (2008).

\bibitem{Grover1997}
 L. K. Grover, {\it Quantum Mechanics Helps in Searching for a Needle in a Haystack}, Phys. Rev. Lett. {\bf 79}, 325 (1997).


\bibitem{Aaronson2011}
S. Aaronson and  A. Arkhipov, The Computational Complexity of Linear Optics. in {\it Proceedings of the 43rd Annual ACM Symposium on Theory of Computing} (Association for Computing Machinery, New York, 2011), pp. 333-342.

\bibitem{Broome2013}
M. A. Broome,  A. Fedrizzi, S. Rahimi-Keshari, J. Dove, S. Aaronson, T. C. Ralph, and A. G. White,  {\it Photonic Boson Sampling in a Tunable Circuit}, Science {\bf 339}, 794-798 (2013).

\bibitem{Spring2013}
J. B. Spring, B. J. Metcalf, P. C. Humphreys, W. S. Kolthammer, X.-M. Jin, M. Barbieri, A. Datta, N. Thomas-Peter, N. K. Langford, D. Kundys, J. C. Gates, B. J. Smith, P. G. R. Smith, and I. A. Walmsley, {\it Boson Sampling on a Photonic Chip}, Science {\bf 339}, 798-801 (2013).

\bibitem{Tillman2013}
 M. Tillmann,	B. Daki\'{c},	R. Heilmann,	S. Nolte,	A. Szameit, and P. Walther, {\it Experimental Boson Sampling}, Nat. Photon. {\bf 7}, 540-544 (2013).

\bibitem{Crespi2013}
A. Crespi, R. Osellame, R. Ramponi, D. J. Brod, E. F. Galv\~{a}o, N. Spagnolo, C. Vitelli, E. Maiorino, P. Mataloni, and F. Sciarrino, {\it Integrated Multimode Interferometers with Arbitrary Designs for Photonic Boson Sampling}, Nat. Photon. {\bf 7}, 545-549 (2013).

\bibitem{LuPan2017Boson}
H. Wang, Y. He,	Y. Li, Z. Su, B. Li, H. Huang, X. Ding, M.-C. Chen, C. Liu, J. Qin, J.-P. Li, Y.-M. He, C. Schneider, M. Kamp, C.-Z. Peng, S. H\"{o}fling, C.-Y. Lu, and J.-W. Pan, {\it High-Efficiency Multiphoton Boson Sampling}, Nat. Photon. {\bf 11}, 361-365 (2017).

\bibitem{Lloyd1996}
S. Lloyd, {\it Universal Quantum Simulators}, Science {\bf 273}, 1073-1078 (1996).

\bibitem{Peng2009}
X.-H. Peng , J. Zhang, J. Du, and D. Suter, {\it Quantum Simulation of a System with Competing Two- and Three-Body Interactions}, Phys. Rev. Lett. {\bf 103}, 140501 (2009).

\bibitem{Peng2010}
X.-H. Peng and D. Suter, {\it Spin Qubits for Quantum Simulations}, Frontiers of Physics In China {\bf 5}, 1 (2010).

\bibitem{AD2010}
 G. A. \'{A}lvarez and D. Suter, {\it NMR Quantum Simulation of Localization Effects Induced by Decoherence}, Phys. Rev. Lett. {\bf 104}, 230403 (2010).

\bibitem{Peng2014}
X.-H. Peng, Z. H. Luo, W. Zheng, S. Kou, D. Suter, and J. Du, {\it Experimental Implementation of Adiabatic Passage between Different Topological Orders}, Phys. Rev. Lett. {\bf 113}, 080404 (2014).

\bibitem{LocdelocTrs2015}
 G. A. \'{A}lvarez, D. Suter, and R. Kaiser, {\it Localization-Delocalization Transition in the Dynamics of Dipolar-Coupled Nuclear Spins}, Science {\bf 349}, 846-848 (2015).

\bibitem{CMC2016}
 M.-C. Chen, D. Wu, Z. Su, X. Cai, X. Wang, T. Yang, L. Li, N. Liu, C. -Y. Lu, and J.-W. Pan, {\it Efficient Measurement of Multiparticle  Entanglement with Embedding Quantum Simulator}, Phys. Rev. Lett. {\bf 116}, 070502 (2016).

 \bibitem{LiOTOC2017}
J. Li, R. Fan, H. Wang, B. Ye, B. Zeng, H. Zhai, X.-H. Peng, and J. Du, {\it Measuring Out-of-Time-Order Correlators on a Nuclear Magnetic Resonance Quantum Simulator}, Phys. Rev. X {\bf 7}, 031011 (2017).

\bibitem{HHL2009}
A. W. Harrow, A. Hassidim, and S. Lloyd, {\it Quantum Algorithm for Linear Systems of Equations}, Phys. Rev. Lett. {\bf 103}, 150502 (2009).

\bibitem{Cai2013}
X. Cai, C. Weedbrook, Z. Su, M.-C. Chen, M. Gu, M. Zhu, L. Li, N. Liu, C.-Y. Lu, and J.-W. Pan, {\it Experimental Quantum Computing to Solve Systems of Linear Equations}, Phys. Rev. Lett. {\bf 110}, 230501 (2013).

 \bibitem{PJ2014}
J. Pan, Y. Cao, X. Yao, Z. Li, C. Ju, H. Chen, X. Peng, S. Kais, and J. Du, {\it Experimental Realization of Quantum Algorithm for Solving Linear Systems of Equations}, Phys. Rev. A {\bf 89}, 022313 (2014).

\bibitem{Barz2014}
S. Barz, I. Kassal, M. Ringbauer, Y. O. Lipp, B. Daki\'{c}, A. Aspuru-Guzik, and P. Walther, {\it A Two-Qubit Photonic Quantum Processor and Its Application to Solving Systems of Linear Equations}, Sci. Rep. {\bf 4}, 6115 (2014).

\bibitem{SupCond2017Lineq}
Y. R. Zheng, C. Song, M.-C. Chen, B. Xia, W. Liu, Q. Guo, L. Zhang, D. Xu, H. Deng, K. Huang, Y. Wu, Z. Yan, D. Zheng, L. Lu, J.-W. Pan, H. Wang, C.-Y. Lu, and X. B. Zhu, {\it Solving Systems of Linear Equations with a Superconducting Quantum Processor}, Phys. Rev. Lett. {\bf 118}, 210504 (2017).

\bibitem{QMLearn2015Lu}
X. Cai, D. Wu, Z. Su, M.-C. Chen, X.-L. Wang, L. Li, N. Liu, C.-Y. Lu, and J.-W. Pan, {\it Entanglement-Based Machine Learning on a Quantum Computer}, Phys. Rev. Lett. {\bf 114}, 110504 (2015).

\bibitem{Li2015}
Z. Li, X. Liu,  N. Xu, and J. Du, {\it Experimental Realization of a Quantum Support Vector Machine}, Phys. Rev. Lett. {\bf 114}, 140504 (2015).

\bibitem{QMLearn2017SupCond}
D. Rist\`{e}, M. Silva, C. Ryan, A. Cross, A. C\'{o}rcoles, J. Smolin, J. Gambetta, J. Chow, and B. Johnson, {\it Demonstration of Quantum Advantage in Machine Learning}, npj Quantum Inf.  {\bf 3}, 16 (2017).

\bibitem{PCA2014}
S. Lloyd,  M. Mohseni,  and P. Rebentrost, {\it Quantum Principal Component Analysis},  Nat. Phys. {\bf 10}, 631 (2014).

\bibitem{WBL2012}
N. Wiebe, D. Braun, and S. Lloyd, {\it Quantum Algorithm for Data Fitting}, Phys. Rev. Lett. {\bf 109}, 050505 (2012).

\bibitem{QImR2016}
F. Yan,  A. M. Iliyasu, and S. E. Venegas-andraca, {\it A Survey of Quantum Image Representations}, Quantum Inf. Process. {\bf 15}, 1-35  (2016).

\bibitem{Venegas2003}
S. E. Venegas-andraca and S. Bose, {\it Storing, Processing and Retrieving an Image Using Quantum Mechanics}, Proc. SPIE Conf. Quantum Inf. Comput. {\bf 5105}, 137-147 (2003).

\bibitem{Le2011}
P. Le, F. Dong, and K. Hirota, {\it A Flexible Representation of Quantum Images for Polynomial Preparation, Image Compression, and Processing Operations},  Quantum Inf. Process. {\bf 10}, 63-84 (2011).

\bibitem{Z2013}
 Y. Zhang, K. Lu, Y. Gao, and M. Wang, {\it NEQR: A Novel Enhanced Quantum Representation of Digital Images}, Quantum Inf. Process. {\bf 12}, 2833-2860 (2013).

\bibitem{QRAM2008}
 V. Giovannetti,  S. Lloyd, and L. Maccone, {\it Quantum Random Access Memory}, Phys. Rev. Lett. {\bf 100}, 160501 (2008).

\bibitem{Grover2002}
L. Grover and T. Rudolph, {\it Creating Superpositions That Correspond to Efficiently Integrable Probability Distributions}, arXiv:quant-ph/0208112.

\bibitem{Soklakov2006}
A. N. Soklakov and R. Schack, {\it Efficient State Preparation for a Register of Quantum Bits},  Phys. Rev. A {\bf 73}, 012307 (2006).

\bibitem{Buhrman2001}
H. Buhrman, R. Cleve, J. Watrous, and R. de Wolf, {\it Quantum Fingerprinting}, Phys. Rev. Lett. {\bf 87}, 167902 (2001).

\bibitem{QCQI2000Book}
M. A. Nielsen and I. L. Chuang, {\it Quantum Computation and Quantum Information} (Cambridge University Press, Cambridge, England, 2000).

\bibitem{Hoyer1997}
P. Hoyer, {\it Efficient Quantum Transforms}, arXiv:quant-ph/9702028.

\bibitem{Fijany1998}
A. Fijany and C. Williams, Quantum Wavelet Transforms: Fast Algorithms and Complete Circuits.  in {\it Proceedings of the 1st NASA International Conference on Quantum Computing and Quantum Communications} (Springer, Berlin, 1998), pp. 10-33.

\bibitem{QFT2001}
Y. S. Weinstein,  M. A. Pravia, E. M. Fortunato,  S. Lloyd, and D. G. Cory,  {\it Implementation of the Quantum Fourier Transform}, Phys. Rev. Lett. {\bf 86}, 1889 (2001).

\bibitem{VCS2010}
 L. M. K. Vandersypen, I. L. Chuang, and D. Suter,``Liquid-State NMR Quantum Computing'' In {\it Encyclopedia of Magnetic Resonance} (John Wiley \& Sons, Ltd., New York, 2010).

 \bibitem{Cory1998}
 D. G. Cory, M. D. Price, and T. F. Havel, {\it Nuclear Magnetic Resonance Spectroscopy: An Experimentally Accessible Paradigm for Quantum Computing}, Physica D: Nonlin. Phenom. {\bf 120}, 82-101 (1998).

\bibitem{NC2011Souza}
 A. M. Souza, J. Zhang, C. A. Ryan, and  R. Laflamme, {\it Experimental Magic State Distillation For Fault-Tolerant Quantum Computing}, Nat. Commun. {\bf 2}, 169 (2011).

\bibitem{Peng2015}
 X.-H. Peng, H. Zhou, B. Wei, J. Cui, J. Du, and R. Liu, {\it Experimental Observation of Lee-Yang Zeros}, Phys. Rev. Lett. {\bf 114}, 010601 (2015).

\bibitem{SpinBook}
M. H. Levitt, {\it Spin Dynamics: Basics of Nuclear Magnetic Resonance}  (John Wiley \& Sons, Ltd., New York, 2008).

\bibitem{Peng2001}
X.-H. Peng, X. Zhu, X. Fang, M. Feng, K. Gao, X. Yang, and M. Liu,  {\it Preparation of Pseudo-Pure States By Line-Selective Pulses in Nuclear Magnetic Resonance}, Chem. Phys. Lett. {\bf 340}, 509-516 (2001).

\bibitem{Grape2005}
N. Khaneja, T. Reiss, C. Kehlet, T. Schulte-Herbr\"{u}ggen, and  S. J. Glaser, {\it Optimal Control of Coupled Spin Dynamics: Design of NMR Pulse Sequences by Gradient Ascent Algorithms}, J. Magn. Reson. {\bf 172}, 296 (2005).

\bibitem{Ryan2008}
 C. A. Ryan,  C. Negrevergne, M. Laforest, E. Knill, and R. Laflamme, {\it Liquid-State Nuclear Magnetic Resonance as a Testbed for Developing Quantum Control Methods}, Phys. Rev. A {\bf 78}, 012328 (2008).

\bibitem{Compiler2016Lijun}
J. Li,  J. Cui, R. Laflamme,  and  X.-H. Peng, {\it Selective-Pulse-Network Compilation on a Liquid-State Nuclear-Magnetic-Resonance System}, Phys. Rev. A {\bf 94}, 032316 (2016).

\bibitem{Chuang1998}
 I. L. Chuang, N. Gershenfeld, M. G. Kubinec, and D. W. Leung, {\it Bulk Quantum Computation with Nuclear Magnetic Resonance: Theory and Experiment}, Proc. R. Soc. A {\bf 454}, 447-467 (1998).

\bibitem{IMED2009}
J. Li and B. Lu, {\it An Adaptive Image Euclidean Distance},  Pattern Recogn. {\bf 42}, 349-257 (2009).

\bibitem{Marr1980Edge}
D. Marr and E. Hildreth, {\it Theory of Edge Detection},  Proc. R. Soc. B {\bf 207}, 187-217 (1980).

\bibitem{Hubel1995}
D. Hubel, {\it Eye, Brain and Vision} (Scientific American Press, New York, 1995).

\bibitem{Qsobel2015}
 Y. Zhang, K. Lu, and Y. Gao, {\it Qsobel: A Novel Quantum Image Edge Extraction Algorithm}, Sci. China Inf. Sci. {\bf 58}, 1-13 (2015).

\bibitem{Ions2011}
C. Ospelkaus, U. Warring, Y. Colombe, K. R. Brown, J. M. Amini,	D. Leibfried, and D. J. Wineland, {\it  Microwave Quantum Logic Gates for Trapped Ions},  Nature (London) {\bf 476}, 181-184 (2011).

\bibitem{Ion2017}
G. Higgins, W. Li, F. Pokorny, C. Zhang, F. Kress, C. Maier, J. Haag, Q. Bodart, I. Lesanovsky, and M. Hennrich, {\it Single Strontium Rydberg Ion Confined in a Paul Trap},  Phys. Rev. X {\bf 7}, 021038 (2017).

\bibitem{SuperCond2015}
J. Kelly,	R. Barends, 	A. G. Fowler,	A. Megrant,	E. Jeffrey,	T. C. White,	D. Sank,	J. Y. Mutus, B. Campbell,	Yu Chen,	Z. Chen,	B. Chiaro,	A. Dunsworth,	I.-C. Hoi,	C. Neill,	P. J. J. O'Malley,	C. Quintana,	P. Roushan,	A. Vainsencher,	J. Wenner,	A. N. Cleland, and John M. Martinis, {\it State Preservation by Repetitive Error Detection in a Superconducting Quantum Circuit}, Nature (London) {\bf 519}, 66-69 (2015).

\bibitem{Kok2007}
 P. Kok, W. J. Munro, K. Nemoto, T. C. Ralph, J. P. Dowling, and G. J. Milburn, {\it Linear Optical Quantum Computing with Photonic  Qubits}, Rev. Mod. Phys. {\bf 79}, 135-174 (2007).

\bibitem{Carolan2015}
J. Carolan, C. Harrold, C. Sparrow, E. Martinlopez, N. J. Russell, J. W. Silverstone, and A. Laing, {\it Universal Linear Optics}, Science {\bf 349}, 711-716 (2015).

 \bibitem{Haar1910}
A. Haar, {\it Zur Theorie der Orthogonalen Funktionensysteme}, Mathematische Annalen  {\bf 69}, 331-371  (1910).

\bibitem{Duality2016}
S. J. Wei and G. L. Long, {\it Duality Quantum Computer and the Efficient Quantum Simulations},  Quantum Inf. Process.  {\bf 15}, 1189-1212 (2016).

\bibitem{Knill2007}
E. Knill, G. Ortiz,  and R. D. Somma, {\it Optimal Quantum Measurements of Expectation Values of Observables}, Phys. Rev. A {\bf 75}, 012328 (2007).

\bibitem{Lloyd2013}
S. Lloyd, M. Mohseni, and P. Rebentrost,  {\it Quantum Algorithms for Supervised and Unsupervised Machine Learning}, arXiv:1307.0411.

\bibitem{Aaronson2009}
S. Aaronson, {\it BQP and the Polynomial Hierarchy}, arXiv:0910.4698.

 \end{thebibliography}
\end{document}